\documentclass{sig-alternate}

\usepackage[ruled,vlined,linesnumbered]{algorithm2e}
\usepackage[utf8]{inputenc}
\usepackage{subfigure}
\usepackage{mathtools}
\usepackage{multicol}
\usepackage{graphicx}
\usepackage{rotating}
\usepackage{mathptmx}
\usepackage{amsfonts}
\usepackage{balance}
\usepackage{amsmath}
\usepackage{amssymb}
\usepackage{wrapfig}
\usepackage{times}
\usepackage{color}
\usepackage{tikz}
\usepackage{url}

\usetikzlibrary{decorations.pathreplacing,shapes}

\newcommand{\eat}[1]{}

\SetKwRepeat{Do}{do}{while}

\title{Parallelized Proximity-Based Query Processing Methods\\ for Road Networks}

\numberofauthors{1} 
\author{
\alignauthor
George Tsatsanifos\\
\affaddr{HomeGrown Research Labs}\\
\affaddr{Omirou 39, Agios Dimitrios,}\\
\affaddr{172 36, Athens, Greece}
}

\begin{document}
\CopyrightYear{2016}

\maketitle
\begin{abstract}
  In this paper, we propose a paradigm for processing in parallel 
  graph joins in road networks. The methodology we present can  
  be used for \emph{distance join processing} among the elements 
  of two disjoint sets $R,S$ of nodes from the road network, with 
  $R \prec S$, and we are in search for the pairs of vertices $(u,v)$, 
  where $u \in R$ and $v \in S$, such that $\textrm{dist} (u,v)\leq 
  \theta$. Another variation of the problem would involve retrieving 
  the $k$ closest pairs $(u,v)$ in the road network with $u \in R$ and 
  $v \in S$, such that $\textrm{dist}(u,v) \leq \textrm{dist}(w,y)$, 
  where $w,y$ do not belong in the result.

  We reckon that this is an extremely useful paradigm with many 
  practical applications. A typical example of usage of our methods 
  would be to find the pairs of restaurants and bars (in that order) 
  from which to select for a night out, that either fall within 
  walking distance for example, or just the $k$ closest pairs, 
  depending on the parameters. Another entirely different scenario 
  would involve finding the points of two distinct trajectories that 
  are within a certain distance predicate, or the $k$ closest such 
  points. For example, we would like to transfer from one train to 
  another a few tones of freight, and hence, we want to minimize the 
  distance we have to cover for moving the cargo from the carrying 
  train to the other. We reckon that this endeavor of ours covers 
  exactly those needs for processing such queries efficiently.

  Moreover, for the specific purposes of this paper, we also propose 
  a novel heuristic graph partitioning scheme. It resembles a recursive 
  bisection method, and is tailored to the requirements of the problem, 
  targeting at establishing well separated partitions, so as to allow 
  computations to be performed simultaneously and independently within 
  each partition, unlike hitherto work that aims at minimizing either 
  the number of edges among different partitions, or the number of nodes 
  thereof.
\end{abstract}
\section{Introduction}
\label{sec:intro}

The majority of the research that has been conducted in the 
area of graph partitioning addresses the $k$-way partitioning 
problem, according to which the vertices of the graph are 
partitioned into $k$ disjoint sets in such a way that the sum 
of weights of the edges whose incident vertices belong to different 
sets is minimized. The problem can be extended so that we have 
an equal number of vertices in each set. Despite the fact these 
works address a very similar problem, they have an entirely different 
motivation than ours. For instance, the efficient execution of many 
parallel algorithms tacitly requires the solution to a graph 
partitioning problem, where vertices represent tasks and edges 
represent data exchanges. Depending on the amount of the computation 
performed by each task, the vertices are assigned to a proportional 
weight. Likewise, the edges are assigned weights that reflect the 
amount of data that need to be exchanged. Hence, by assigning to 
each processor tasks whose computational cost is almost the same 
and by minimizing the communication overhead that corresponds to the 
edge-cut the overall response time is optimized. The same problem 
definition is used for the fast sparse matrix-vector multiplication, 
when representing graphs as sparse matrices. Also, to compute a 
fill-reducing ordering that leads to a high degree of concurrency 
in the factorization phase of solving sparse system.

Our work has a similar optimization incentive though for processing 
spatial graphs and a different formulation of the problem from a 
different prospective needed to be provided in order to obtain a 
similar effect, capturing the essential differences in the nature 
of the problem and an entirely different approach had to be followed. 
Before that we should define the types of queries we are interested 
in. Then, it will be easier to describe the specifications of a graph 
partitioning scheme that would boost the performance for those types 
of queries. More formally, among the elements of two disjoint sets 
$R,S$ of nodes from a road network, with $R \prec S$, we are in 
search for the pairs of vertices $(u,v)$, where $u \in R$ and $v \in 
S$, such that $\textrm{dist} (u,v)\leq \theta$. According to a 
variation of the problem, we are in search for the $k$ closest pairs 
$(u,v)$ in the road network with $u \in R$ and $v \in S$, such that 
$\textrm{dist} (u,v) \leq \textrm{dist}(w,y)$, where $w,y$ do not 
belong in the result. Most importantly, all distances are computed over 
the road network and do not relate to the respective distances in the 
Euclidean space. Albeit a seemingly small and intricate detail, it 
completely changes the techniques that can be employed and the nature 
of the methods to be developed. Moreover, for processing those types 
of queries, we propose a novel heuristic graph partitioning scheme 
which resembles a recursive bisection method and targets at establishing 
well separated partitions, so as to allow computations to be performed 
simultaneously and independently within each partition, unlike hitherto 
work that aims at minimizing either the number of edges among different 
partitions, or the number of nodes thereof. To the best of our knowledge, 
this paper constitutes the very first effort to address the problem of 
graph partitioning in the context of distance join and closest pairs 
computations. More importantly, we are in position of speeding up the 
processing of such queries by parallelizing the process as much as possible 
by processing each partition of the graph independently and also minimizing 
the cost involved in merging the partial results. Last but not least, we 
can easily extend our methods to operate over mobile objects as well, by 
specifying appropriately the distances of each from the two ends of the 
edge it moves on. This information can either be incorporated within the 
graph (but carefully maintained), or kept updated in a different spatial 
data structure for faster retrieval. Those objects can naturally belong in 
either $R$ or $S$ without any loss, and search can be initiated from them 
just as efficiently with a minor twist at the beginning of processing each 
such query.

The remainder of the paper is organized as following: In 
Section \ref{sec:related}, we discuss the relevant literature, 
in Section \ref{sec:partitioning}, we propose a heuristic 
partitioning scheme, in Section \ref{sec:processing}, we 
present our paradigm for processing such queries, in Section 
\ref{sec:exp} we evaluate our methods, and finally, in Section 
\ref{sec:concl} we conclude and summarize our contributions.

\section{Related Work}
\label{sec:related}

In this section we present the relevant literature regarding graph 
partitioning and schemes for query processing in road networks.

\subsection{Graph Partitioning}

Arguably the most successful class of partitioning algorithms 
are the multilevel graph partitioning schemes. Those algorithms 
try to reduce the graph by collapsing vertices and edges, 
partitions the smaller graph, and then uncoarsens it to the 
most preferable outcome. In this context, the authors in 
\cite{HendricksonL95} present a multi-level algorithm in which 
the graph is approximated by a sequence of increasingly smaller 
graphs that are partitioned using a spectral method. A variant 
of the Kernighan-Lin algorithm \cite{Kernighan70} is applied 
periodically to refine the partition. More specifically, the KL 
algorithm examines all vertices with respect to the improvement 
that they bring to the partitioning. This corresponds to the net 
reduction in the weight of cut edges that would result from 
switching a vertex to a different set. Vertices with the opposite 
effect are also considered since several moves that reduce the 
partition quality may lead to later moves that compensate 
for the initial regression. To this feature can be attributed 
the ability of the algorithm to climb out of local minima.
Of course, with each assignment, the ranks of all adjacent 
vertices are updated accordingly. Eventually, the best 
partition that is encountered in this sequence is selected 
with each iteration. An improved linear time implementation 
of this is proposed in \cite{FiducciaM82}. A meticulous study 
on the characteristics of these methods can be found in 
\cite{KarypisK98a} together with improvements in terms of 
response time that rely on a finer refinement heuristic, and 
in quality due to the policy they adopt during contracting 
nodes that preserves the properties of the original graph. 
Many strategies are investigated for coarsening, including 
selecting random nodes, tracing cliques, and others. In 
addition a greedy graph growing partitioning algorithm is 
proposed that resembles breadth-first search. A parallel 
implementation of this is proposed in \cite{KarypisK99} that 
relies on graph coloring. It is shown that these works provide 
better partitions than spectral methods at lower cost for a 
variety of finite element problems.

In retrospect, spectral methods are expensive since they require 
the computation of the eigenvector corresponding to the second 
smallest eigenvalue. Geometric partitioning algorithms tend to 
be fast but often yield partitions that are worse that those 
obtained by spectral methods. However, they are applicable only 
if coordinates are available for the vertices of the graph, a 
feature that is not always available.

More recently, in \cite{StantonK12} a streaming model is adopted 
for graph partitioning, which can be used in combination with a 
number of different heuristics that can be employed under different 
circumstances. Different criteria are used, like trying to balance 
the size of each partition, the cardinality of the cut edge set, 
hashing, etc. Furthermore, distributed technologies like map-reduce 
and peer-to-peer technologies have lavished attention for the last 
years. Towards this trend, the authors in \cite{RahimianPGJH13} 
propose a distributed scheme for for solving the balanced $k$-way 
graph partitioning problem. The intuition of their method is quite 
simple. They assign each vertex to a separate partition with a 
probability analogous to the desired size of the partition. Then, 
each node iteratively selects another node from either its neighbors 
or a random sample, and examines the pair-wise benefit of a color 
exchange. In a simulated annealing fashion, if the color exchange 
results in a more preferable partitioning, then the two nodes swap 
their colors. Hence, non-preferable outcomes are also acceptable 
since they can lead to a better partitioning than a greedy heuristic 
approach would.

Notably, all of the above partitioning schemes have an entirely 
different motivation and used under different circumstances. Only 
a dearth of work can be found regarding partitioning schemes for 
processing road networks. The most similar work can be found in 
\cite{proximity}, where the authors propose using certain paths 
of the graph in order to partition it. Under this setting, the 
boundary of a (sub)graph corresponds to the sequence of edges 
that form a cycle bounding a planar embedding of it. In particular, 
those paths are chosen in such a way that with each cut the graph 
is divided recursively into partitions of approximately equal size. 
As long as the end nodes of each cut are selected from the boundary 
of the original graph (not from nodes on a cut), no cuts generated 
in this manner intersect with each other. In other words, we end 
up with a set of ``parallel'' non-intersecting cuts. This approach 
works well for city maps where most of the streets are parallel 
forming a grid-like pattern. On the downside, processing the studied 
query types involves a computational overhead owing to the fact that 
the different partitions are too close to each other as there is no 
mechanism to guarantee that the partitions are enough far apart from 
each other in order to prevent that. Last but not least, our scheme 
was designed under the spectrum of performing parallel operations.

\subsection{Graph Processing}

One of the earliest efforts in this field can be traced in 
the seminal paper \cite{papadias} that addresses processing 
spatial queries in road networks efficiently using lower 
bounds that rely on the Euclidean distance between the nodes.
In particular, they present algorithms for (i) range queries, 
(ii) nearest neighbor queries, (iii) closest pairs retrieval, 
and (iv) $\theta$-joins according to network distance. For 
each query type they present two algorithms, for relying on 
Euclidean space before operating over the road network, and 
vice versa. However, the Euclidean distance is a rather coarse 
metric that not always associates with the weights of the edges. 
If, for instance, the edge cost is defined as the expected 
travel time, the Euclidean distance cannot confine the search 
space (unless we make additional assumptions, such as maximum 
speed). In addition, the lack of an upper bound to guide the 
network expansion leaves a margin for improvements.

The authors in \cite{KriegelKKRS07} present a scheme for processing 
range queries and $k$NN queries in road networks fast by relying on 
graph embedding, a structure that is built during a preprocessing 
stage to compute the distances of all nodes from specific landmarks. 
In particular, their methods rely on the notion of a set of reference 
nodes, from which a subset of distances is kept from each node. Notably, 
they showed in their evaluation that keeping just the five such closest 
reference nodes (and thus the set of global reference nodes is allowed 
to grow larger) is enough to accomplish stellar performance. In addition, 
suitable functions are proposed so that a lower and an upper bound can 
be established that can be used in a filter/refinement query processing 
architecture. Also, those can be exploited in an $A^*$ implementation to 
outperform any approach relying on Euclidean distances. An extension of 
that work can be found in \cite{KriegelKRS08} where the authors propose a 
hierarchical embedding that scales well to large traffic networks. A 
number of layers of reference nodes forming a complete graph by necessity 
is kept, and by processing each time a different partition rather than 
just the flat embedding at the bottom layer, performance ameliorates.
Albeit a successful approach for processing range and $k$NN queries, 
the query types studied in this paper require different tools for 
performing searches simultaneously in different parts of the graph, 
while thoroughly coordinating the radius of each local search according 
to the other partial results, so that it does not exceed any unnecessary 
levels to render our approach inefficient. Even though, a partition 
hierarchy is fundamental for our processing techniques too, the one 
proposed in the context of this work is built differently taking completely 
different parameters into consideration, and we process it in a bottom-up 
fashion instead, starting from the relevant leaf nodes, after a short 
top-down preparation phase to find those and plan the execution of the 
query throughout the levels of the hierarchy. We reckon that no other form 
of indexing is required for the query types we address in this paper.
Even for geographically constrained queries of the kind, we can easily 
restrict processing so as to involve only the partitions that satisfy 
the constraints by a simple comparison with the MBR of the partition 
and prevent from processing the rest. Other than that, the exact same 
processing steps are undertaken.

In \cite{SankaranarayananS10}, the authors approximate distances with 
close values relying on the observation that the distance distortion 
(i.e., the ratio of the network distance to the spatial distance between 
two vertices) decreases as the separation between the vertices increases.
In other words, large distortions occur at small spatial distances, 
while the distortion quickly reduces to smaller values as he sources 
and destinations get farther away. They use bounds from above and below 
for the distance between two vertices which are used to select distance 
approximations within an $\epsilon$ factor.
In another line of work, query processing in \cite{proximity} relies on 
proximity relations differs from the range query and nearest neighbor query, 
as it operates over, constrain, and monitor the constellation among sets of 
moving objects. The authors in \cite{proximity} consider scenarios in road 
networks for monitoring spatial relations and the efficient evaluation of 
queries for large numbers of concurrently moving objects over the road 
network.

\section{Heuristic Graph Partitioning}
\label{sec:partitioning}

The method we present in this section takes as input a graph 
partition and splits it into two parts, in such a way that 
the two derived graph partitions are well separated, in a 
sense that we want to distance them as far from each other 
as possible. The two derived partitions may be connected with 
each other with edges that have their two ends in different 
clusters, which from now on we will refer to as \emph{cross-edges}. 
We will call the vertices that are adjacent to those edges 
\emph{border nodes}. Since this is a NP-hard problem of 
combinatorial nature, we rely on greedy heuristics. and 
propose an approximation scheme on how to solve this problem.
Most importantly, we can perform this way the required computations 
\emph{simultaneously} and \emph{independently} within each partition, 
so as to come up with as many partial and local results, which of 
course need to be combined accordingly at a higher level. Apart from 
that, separate computations need to be performed starting from the 
edges that run through the different partitions. Thereby, the further 
apart the partitions are, and the stricter the thresholds arising from 
the local results, the less computations need to be performed and the 
performance of our paradigm ameliorates. As a matter of fact, the 
less computations are required to derive results among different 
partitions, the greater the throughput and the performance gain 
because of the parallelism we enforce with our methods.

\begin{algorithm}[htbp]
  create an empty left cluster\;
  create an empty right cluster\;
  insert leftvertex into the left cluster\;
  insert rightvertex into the right cluster\;

  create a heap for exploring around the left cluster\;
  create a heap for exploring around the right cluster\;

  \ForEach{edge adjacent to leftvertex}{
    insert it to the heap for the left cluster\;
  }

  \ForEach{edge adjacent to rightvertex}{
    insert it to the heap for the right cluster\;
  }

  create an empty left child partition\;
  create an empty right child partition\;

  \ForEach{cross-edge in cross-edges}{
      insert into left child cross-edges\;
      insert into right child cross-edges\;
  }

  \While{either of the heaps is not empty}{
    \If{right heap is empty or left heap top item better than right heap top item}{
      \If{left heap front end of top item not in left cluster}{
	insert front end from top of heap into left cluster\;
	\ForEach{edge adjacent to front end of the top of heap associated with left cluster}{
	  \If{edge not a cross-edge}{
	    \If{front end of edge not in any cluster}{
	      insert edge into heap for left cluster;
	    }\Else{
	      \If{front end of edge in right cluster}{
	          insert edge into left child partition cross-edges set\;
	          set separationdegree as needed;
	      }
	    }
	  }
	}
      }
      remove top item of heap associated with left cluster\;
    }\Else{
      \If{right heap top item front end not in right cluster}{
	insert front end from top of heap into right cluster\;
	\ForEach{edge adjacent to front end of the top of heap associated with right cluster}{
	  \If{edge not a cross-edge}{
	    \If{front end of edge not in any cluster}{
	      insert edge into heap for right cluster;
	    }\Else{
	      \If{front end of edge in right cluster}{
	          insert edge into right child partition cross-edges set\;
	          set separationdegree as needed;
	      }
	    }
	  }
	}
      }
    }
    remove top item of heap associated with right cluster\;
  }
  \caption{GraphPartition.partition (Vertex leftvertex, Vertex rightvertex)}
  \label{algo:partition}
\end{algorithm}

Algorithm \ref{algo:partition} takes as input two vertices of the 
road network in order to perform the partitioning process. From 
those two vertices we initiate two graph traversals in a best-first 
search fashion, as in a bidirectional expansion, and each time we 
insert the vertices we encounter from either of the two horizons 
into the appropriate cluster. The nodes in those two clusters will 
constitute the nodes of the respective derived partitions. With 
each iteration we choose to expand the cluster with the shortest 
edge on its horizon. Thereby, we can prove that if there is just 
one edge left at the end of the process to serve as a cross-edge, 
it will be necessarily the longest one. Arguably, when the set of 
cross-edges comprises of more edges, it consists of very long 
edges, as well, among which is the longest edge of course. Most 
importantly, we are interested in maximizing the minimum distance 
between any two partitions of the whole constellation of partitions.

\begin{algorithm}[htbp]
  \If{gp is a leaf partition}{
    process locally partition gp for $R,S$ to get $k$ items within $\theta$\; 
  }\Else{
    create an empty thread stack\;
    create a heap for running threads\;
    
    create a thread instance for the root partition\;
    insert new thread into the thread stack\;

    \While{thread stack is not empty}{
      get next thread to examine\;

      \If{thread is ready to run}{
	\If{number of running threads exceeds parallelism}{
	  wait for at least one thread to finish\;
	}
	initiate execution of examined thread\;
      }\Else{
	  \If{thread not associated with leaf partition}{
	    initialize thread associated with left partition\;
	    insert it into the thread stack\;
	    initialize thread associated with right partition\;
	    insert it into the thread stack\;
	    mark examined thread as expanded\;
	  }
      }
    }
  }
  \caption{closestPairs (GraphPartition gp,$R$,$S$,$k$,$\theta$)}
  \label{algo:closestpairs}
\end{algorithm}

Furthermore, we provide a smoothing parameter $\alpha$ 
that takes values in $[0,1]$, according to which we can 
allow taking into consideration the relative populations 
of the graph partitions, upon the decision of which 
partition to chose for a given vertex. Thereby, we can 
use for clustering the updated weights for the two 
clusters, each time normalizing the weight of the 
edge at the top of respective heap as following:

\begin{eqnarray}
 w_1' = \frac{|V_1|}{|V_1|+\alpha |V_2|} w_1 \\
 w_2' = \frac{|V_2|}{\alpha |V_1|+|V_2|} w_2
\end{eqnarray}

where $|V_1|$ the current population of the first partition, 
and $|V_2|$ the population of the second. Of course, $|V_1|+
|V_2|$ the population of the parent graph partition. Subsequently, 
for $\alpha=0$, we have that $w_1'=w_1$ and $w_2'=w_2$, whereas 
for $\alpha=1$, we take $w_1'=\frac{|V_1|}{|V_1|+|V_2|} w_1$ 
and $w_2'=\frac{|V_2|}{|V_1|+|V_2|} w_2$. 

This smoothing parameter when configured appropriately can 
prevent the formation of partitions with immense population 
discrepancies. The selection of the optimal value for $\alpha$ 
is not straightforward and is affected by a large number of 
parameters. To elaborate, we reckon that it is important 
creating partitions that are located as distant as possible 
from each other, as this is the best way to minimize the 
required computations that takes place in the above layers 
when merging the partial results. Nevertheless, this sophistication 
is far from unnecessary as extravagant population imbalances, 
would not allow reciprocate the load at the bottom levels of 
the hierarchy. For instance, a thread processing the left 
child partition of a node could terminate unexpectedly earlier 
than the thread processing the right. Unless the thread 
manager/scheduler detects those occurrences and reacts by 
joining the terminated thread so as to allow its resources 
to be allocated by a thread processing another partition, 
those processing resources would soon be idled. And still, 
even though the load at the higher levels would be diminished, 
we would have to wait until the computations in both partitions 
end before we start processing their results further at the 
higher levels. Therewith, we soon recognized the need to 
mitigate and blunt those imbalances with a configurable 
parameter so as to allow for compromising the trade-off 
between the two extreme cases.

\newpage
\section{Parallel Join Processing}
\label{sec:processing}

In this section, we present a scheme for retrieving the $k$ closest pairs 
$(u,v)$ of a road network, with $u \in R$ and $v \in S$. Nevertheless, we 
have formalized the interface of our paradigm in such a way that we can 
limit processing withing a predefined distance $\theta$ from any point of 
$R$, as we do not allow search for matches to extend beyond that distance 
threshold. Thereby, by setting the threshold parameter $\theta$ appropriately, 
and by not defining a restrictive number of results $k$, we can perform a 
distance join operation according to the specific distance predicate $\theta$. 
And withal, the versatility and usefulness of our paradigm becomes clear. 
More formally, this operation given two sets of points to be matched $R$ 
and $S$, and a distance predicate $\theta$, would return a result comprising 
the pairs of points $(u,v)$, where $u \in R$ and $v \in S$, such that $dist(u,v) 
\leq \theta$.

The main intuition behind our paradigm is to process each partition 
independently using Algorithm \ref{algo:localpairs} so as to retrieve 
the $k$ local closest pairs, and then, process their associated 
cross-edges in Algorithm \ref{algo:combinepairs}, in such a way that 
they are expanded in Algorithm \ref{algo:expandcrossedge} only as much 
as it is required for the results to be updated accordingly. We also 
incorporate a distance predicate $\theta$, so as not to allow matching 
of nodes beyond that threshold and by not defining an expected result 
size, a different query type is also supported. Then, the derived results 
are merged together to form a single set by removing each time the worst 
ranked elements until just $k$ items can be found. 

More importantly, we can parallelize the process, and build a structure 
for processing the graph partitions of the hierarchy in such a way that 
each time we can process up to a number of partitions simultaneously 
according to the desired degree of parallelism. For parallel processing, 
we create recursively a pool of threads with Algorithm \ref{algo:closestpairs}, 
according to the partition hierarchy, and we execute those threads in 
reverse with the help of the stack we create in Alg.~\ref{algo:closestpairs}, 
line 4. First, we insert in lines 6, 7 the thread that is associated with 
processing the top partition, in other words, for merging the results from 
the lower partitioning layers, which of course is run last. In lines 9--13, 
we examine the next thread from the stack and check if it is ready to run, 
and by this, we mean that the threads associated with the partitions of the 
lower layer have been executed and their results are made available to the 
upper layer. Otherwise, we create for the thread pool the threads that correspond 
to the next layer for further processing. Additionally, an auxiliary priority 
queue is used to keep track of the running threads, and manage them appropriately 
in lines 5, 12. The executed threads are prioritized according to 
their distance from the closest leaf they subsume. The motivation behind this 
is that since we have initiated the execution of threads that are either 
running or are waiting for the results of other threads, we should first 
gather the results of the threads associated with the lower levels of the 
hierarchy, and also make available the resources they had allocated as part 
of their local processing tasks. Otherwise, by waiting to free the resources 
of the waiting threads, we waste significant time and the level of parallelism 
of the applications drops dramatically. Therefore, we choose to join first 
the threads that are associated with the partitions at the bottom levels, since 
they are most likely to be involved with processing over the part of the graph 
they represent, rather than combining the results of subsumed layers. 

\begin{algorithm}[htbp]
  \While{elements from both local results more than $k$}{
    \If{worst item from the left better than worst from the right}{
      remove worst item from the right\;
    }\Else{
      remove worst item from the left\;
    }
  }

  insert the $k$ remaining items into a single result-set\;

  \ForEach{cross-edge from the left child partition to right} {
    \If{weight of cross-edge greater than all $k$ result items}{
      \textbf{break}\;
    }
    read global threshold $\Theta$ atomically\;
    \ForEach{match \textbf{in} expandCrossEdge (gp, crossedge, leftbordernodes.get(crossedge.from), $R$, $S$, $k$, $\Theta$)}{
      update the result inserting new match\;
      read global threshold $\Theta$ atomically\;
      if required, update global threshold $\Theta$\;
    }
  }

  \ForEach{cross-edge from the right child partition to left} {
    \If{weight of cross-edge greater than all $k$ result items}{
      \textbf{break}\;
    }
    read global threshold $\Theta$ atomically\;
    \ForEach{match \textbf{in} expandCrossEdge (gp, crossedge, leftbordernodes.get(crossedge.from), $R$, $S$, $k$, $\Theta$)}{
      update the result inserting new match\;
      read global threshold $\Theta$ atomically\;
      if required, update global threshold $\Theta$\;
    }
  }
  \Return result;
  \caption{combinePairs (GraphPartition gp, MaxHeap leftpairs, MaxHeap rightpairs, Set $R$, Set $S$, int $k$)}
  \label{algo:combinepairs}
\end{algorithm}

Since we already know in advance the desired level of parallelism $P$, we 
can tweak scheduling in such a way that the overall throughput is maximized. In 
practice, we do not want to process locally just the leaf nodes of the hierarchy, 
but instead, we want to stop traversing through the partitions at the levels of 
the hierarchy that would ensure us that the degree of parallelism is maximized.
This is accomplished by carefully selecting the appropriate degree of granularity 
the we process the graph partitions, neither finer, nor rougher.

Algorithm \ref{algo:closestpairs} retrieves the $k$ local closest pairs $(u,v)$, 
where $u \in R$, $v \in S$ and $dist(u,v) \leq dist(w,y), \forall w \in R, y \in 
S$. Starting from each element $w \in R$, we perform best-first search, and each 
time, we encounter the next adjacent node $y$, we test whether it belongs in $S$ 
or not. If so, we compare it against the up to $k$ best matched pairs that we have 
retrieved so far, and if it is better than the worst of them, then, we remove that 
and insert the better element, so as to have $k$ items at all times. This is used 
in Algorithm \ref{algo:combinepairs} that takes as input a partition $gp$, two sets 
of points $R, S$ to be matched, and the expected result-size $k$. In lines 1--5, we 
remove all redundant heap elements so as to keep only the $k$ closest pairs from 
both heaps. The remaining elements are merged together appropriately into a single 
heap in line 6. Next, we examine in lines 7--22 whether there are any additional 
pairs that run through the cross-edges of both partitions, left child first in lines 
7--14 and right child in lines 15--22, following the same procedure for both. We 
are allowed to stop early in lines 8--9 and 16--17, since we examine the respective 
cross-edges in ascending order, we are hence in position of ascertaining whether 
it is futile to continue for no better pair can be found to update the result. 
Moreover, in lines 10, 12--14, 18 and 20--22, we access for read or write a global 
distance threshold that is used to limit local search within specific bounds that 
are dictated by the best $k$-th element among all partial results. If each parallel 
thread operated independently then that threshold would be much looser for every 
operation until the final merging at the root of the hierarchy. In practice, this 
is translated into more expensive search operation that overlap with additional 
partitions, something we could easily avoid with a global threshold variable and 
the appropriate read and write locks that suit that kind of concurrent operations.

\begin{algorithm}[htbp]
        initialize an empty result\;
        create an empty heap to store examined paths\;
        insert the input cross-edge to the heap\;

        \While{there are more paths to be expanded}{
          examine the next shortest path from the heap\;
          \If{path cost is greater than threshold $\theta$}{\Return matches\;}
          \If{adjacent node is contained in $S$}{
            \ForEach{partial path in insideroutes}{
                \If{result can be improved or has less than $k$ items}{
                    update result with examined path\;
                }
            }
          }
        }

        \ForEach{edge from front end of the examined path}{
            \If{it is undiscovered and not a cross-edge}{
                extend the examined path with the edge\; 
                insert the new path into the heap\;
            }
        }
        \Return result;
        \caption{expandCrossEdge 
        (GraphPartition gp,Edge crossedge,Map insideroutes, 
        Set $R$,Set $S$,int $k$, double $\theta$)}
        \label{algo:expandcrossedge}
\end{algorithm}

Furthermore, we invoke Algorithm \ref{algo:expandcrossedge} in lines 11 and 19 for 
expanding in tandem the cross-edges that separate the processed partition from its 
sibling. Algorithm \ref{algo:expandcrossedge} also takes as input a threshold parameter 
$\theta$ according to the distance of the worst pair of the derived result. This 
method enacts a search for all pairs that contain the examined cross-edge and their 
distance does not surpass the given local threshold. In lines 6--7 we terminate the 
execution of the method for the next retrieved pair exceeds the given threshold.
In particular, each retrieved pair consists of a vertex from $R$ in the partition 
at the back end of the cross-edge, and another vertex from $S$ (line 8) in the 
partition at the front end of it. Moreover, we are in position of exploiting the 
work we have done in the previous stage during processing locally each partition.
More specifically, whenever we encountered a cross-edge as part of local processing, 
we keep track of the distance from any point $u$ within the previously processed lower 
level partition such that $u \in R$ to the back end of the cross-edge when integrating 
the elements of the various local results into one at a higher level. This way, the 
paths are expanded only from one side here, since the back side of the cross-edge has 
already been previously expanded from an element of set $R$. This happens in lines 4 
and 5 that operate on the inside routes map structure given as input. Of course, the 
returned result would contain no more than $k$ elements.

\section{Experimental Evaluation}
\label{sec:exp}

In this section, we evaluate the performance of our methods 
for various scenarios and configurations of the partition 
hierarchy.

\subsection{Setting}

A variety of different parameters has been investigated to 
illustrate the efficiency of our method: (1) The value of 
the smoothing parameter $\alpha$ used when constructing the 
graph partitions, so as to smooth between the effect of the 
size of the partition, and their pairwise distances. (2) The 
degree of parallelism expressed in the number of running threads. 
(3) The cardinality of the set at the left of the join operation. 
This is expressed as a percentage over the total number of vertices 
in the road network. (4) The cardinality of the set at the 
right of the join operation. (5) The expected result-size for 
$k$ closest pairs. 
Table \ref{table:parameters} presents all parameters along with 
the range of values that they take and their default values.

\begin{table}[htb]
\centering
\begin{tabular}{c c c}
\hline
\textbf{Parameter} & \textbf{Range} & \textbf{Default} \\
\hline
smoothing $\alpha$ & $0.0,0.25,0.50,0.75,1.0$           & $0.0$  \\
parallelism        & $2,4,6,8,10,12,14$                 & $8$   \\
result-size $k$    & $20,40,60,80,100,120,140$          & $80$  \\
$R$ size           & $2\%,4\%,6\%,8\%,10\%,12\%,14\%$   & $8\%$ \\
$S$ size           & $2\%,4\%,6\%,8\%,10\%,12\%,14\%$   & $8\%$ \\
\hline
\end{tabular}
\caption{System parameters.}
\label{table:parameters}
\end{table}

Furthermore, we use two real datasets from \cite{lifeifei}, 
we shall henceforth refer to as NA and SF,  which 
correspond to the interstate network of whole North 
America containing 175,813 nodes and 179,179 edges, 
and San Francisco containing 174,956 nodes and 223,001 
edges. Upon each of these datasets we created the sets 
$R,S$ of vertices to be joined together in such a way 
that correspond to certain proportions of the dataset 
according to the parameters we defined earlier. All 
queries were executed in a 16-core AMD Opteron processor 
tweaked at 2.3 GHz running a server Linux distribution.

\subsection{Results}

\begin{figure}[!htbp]
    \centering
    \subfigure[smoothing parameter]{
    \includegraphics[width=.3\textwidth]{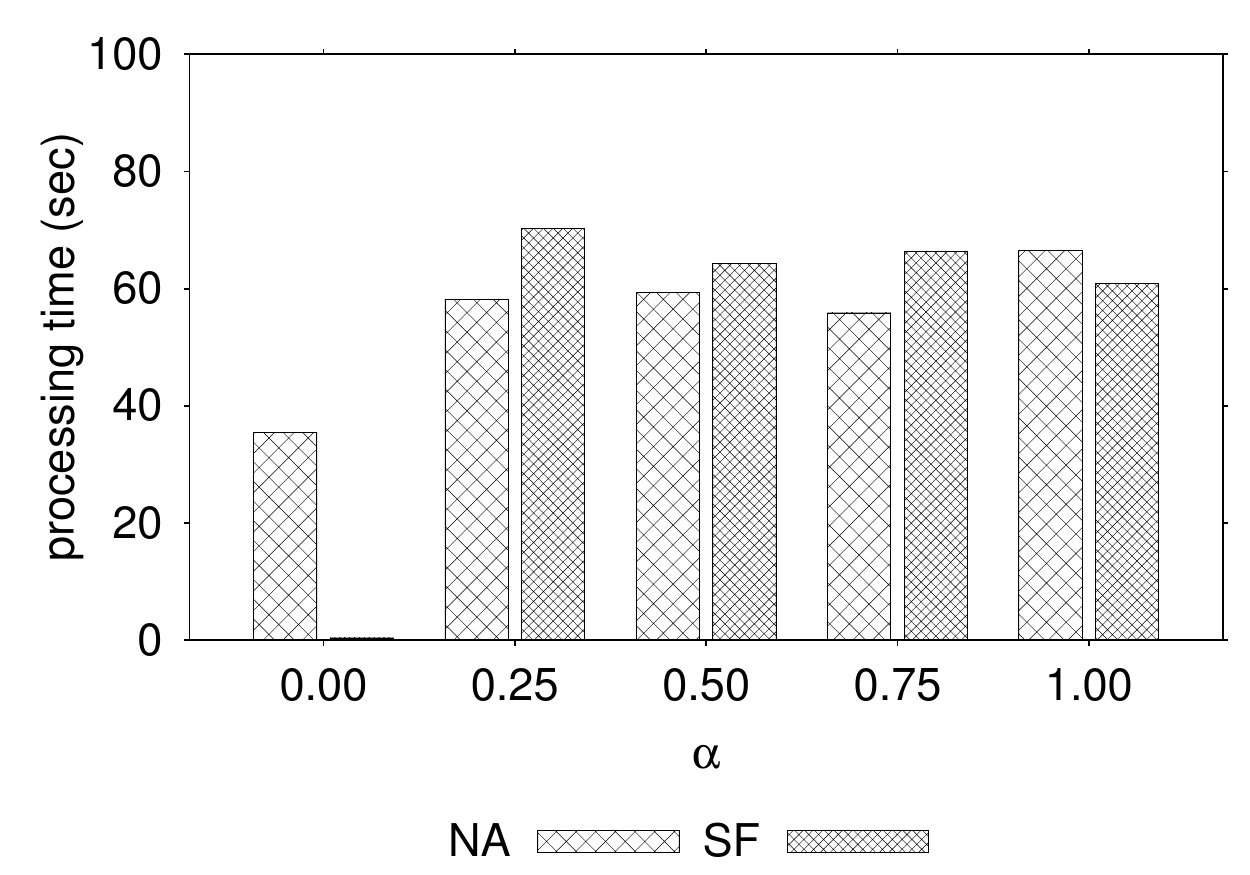}
    \label{fig:results:system:alpha}
    }

    \vspace{-8pt}
    \subfigure[degree of parallelism]{
    \includegraphics[width=.3\textwidth]{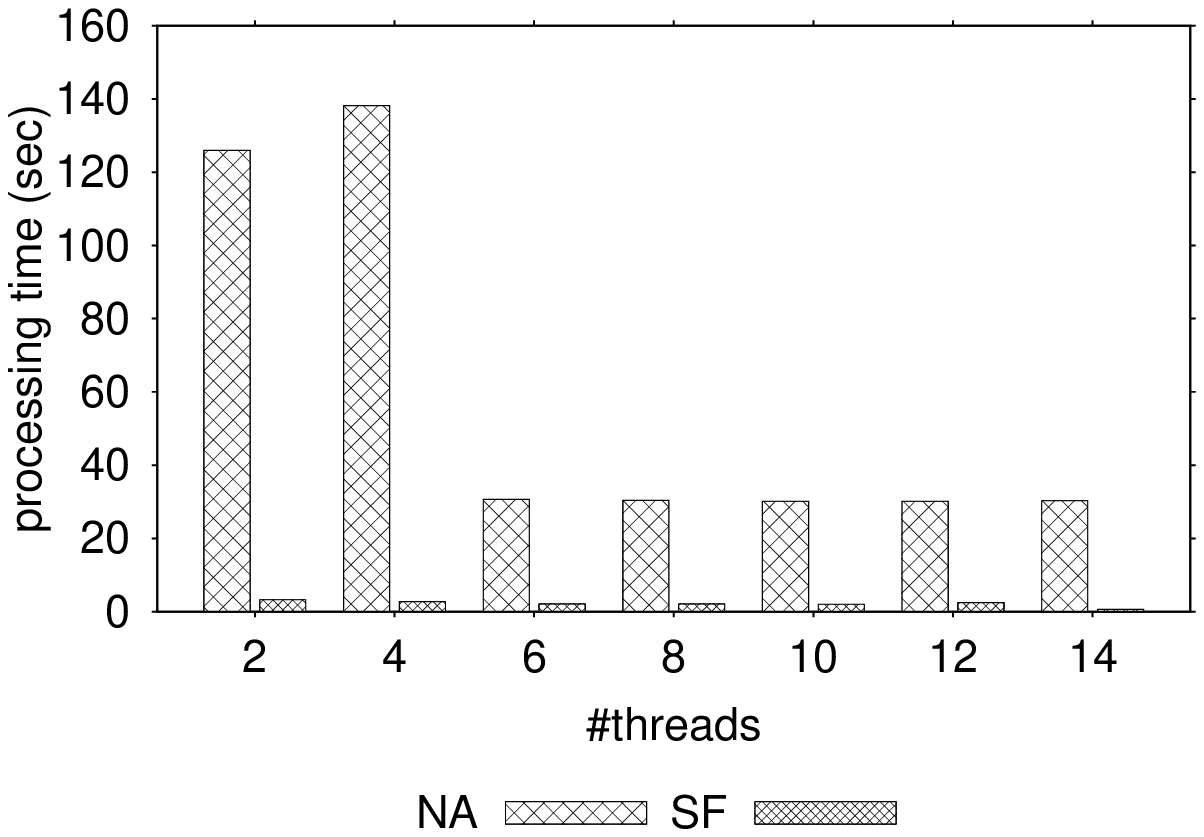}
    \label{fig:results:system:parallelism}
    }

    \vspace{-5pt}
    \caption{System configurations of the partitioning scheme.}
    \label{fig:results:system}
    \subfigure[result-size $k$]{
    \includegraphics[width=.3\textwidth]{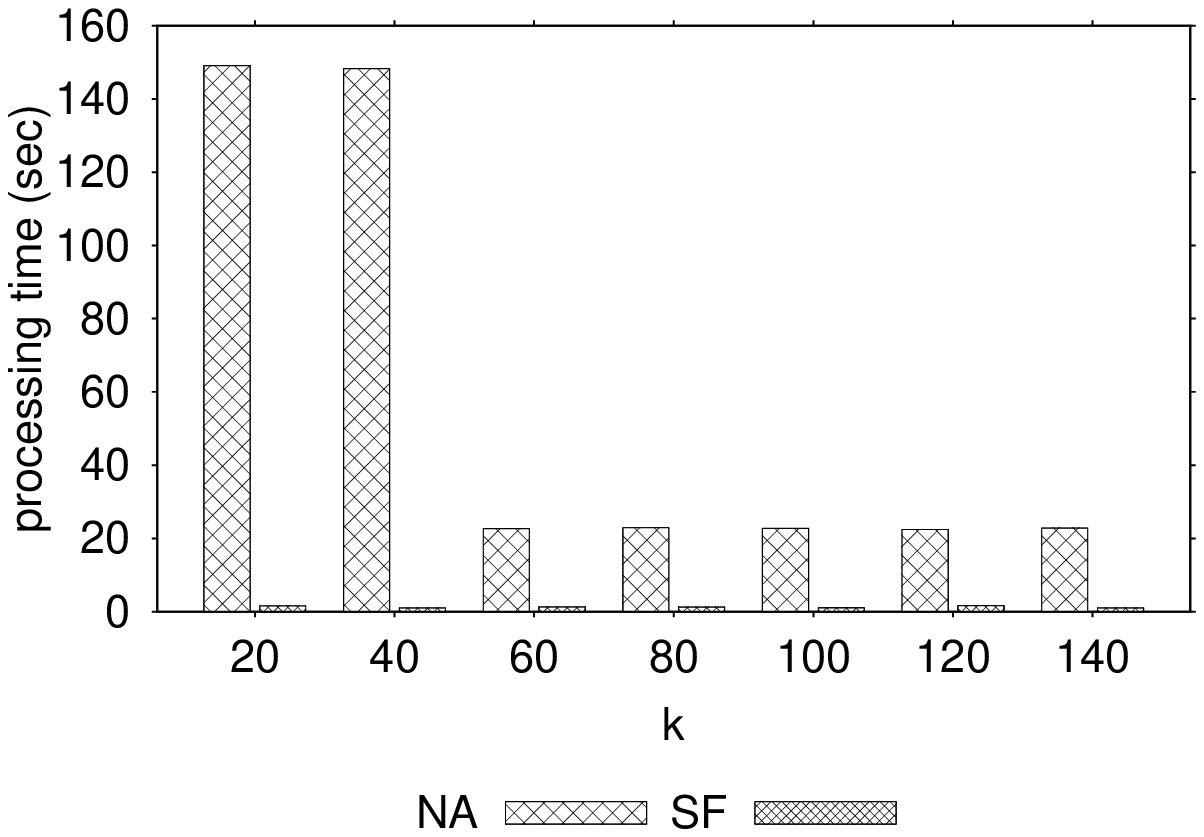}
    \label{fig:results:query:kappa}
    }

    \vspace{-8pt}
    \subfigure[$R$ cardinality]{
    \includegraphics[width=.3\textwidth]{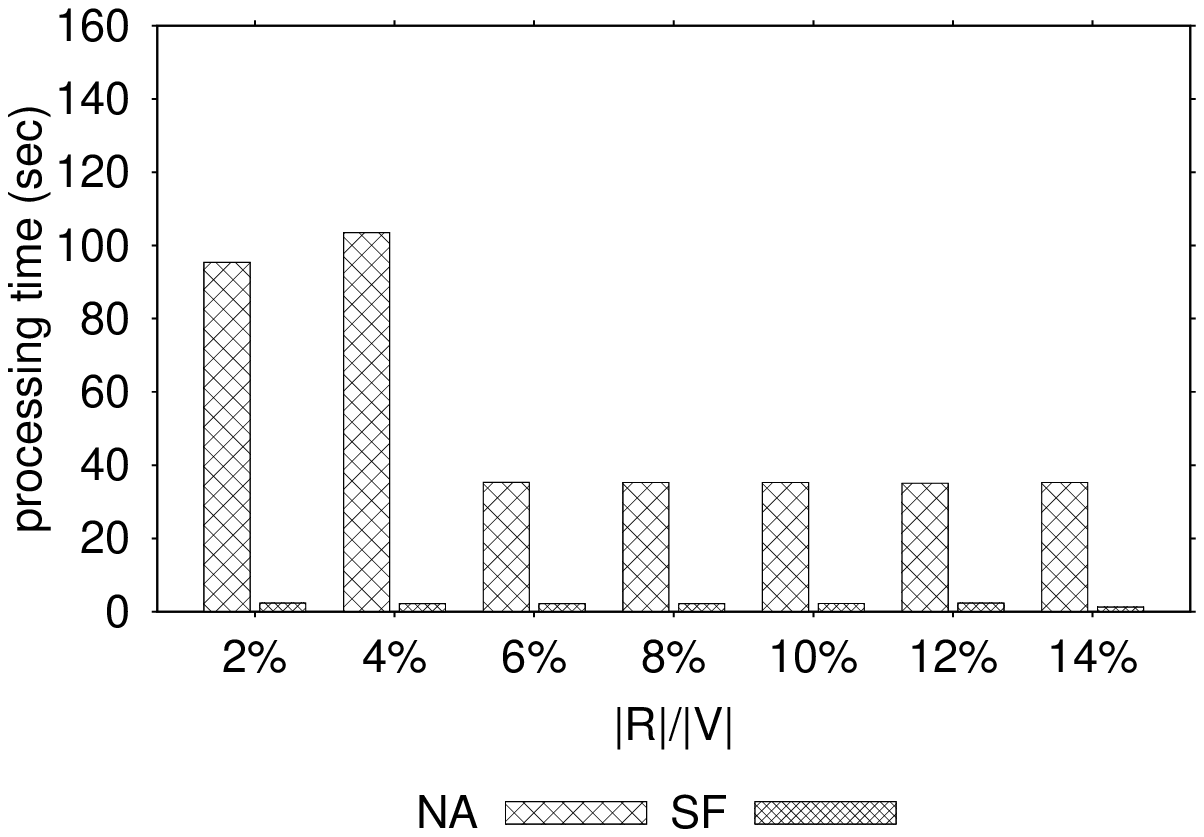}
    \label{fig:results:query:rsize}
    }

    \vspace{-8pt}
    \subfigure[$S$ cardinality]{
    \includegraphics[width=.3\textwidth]{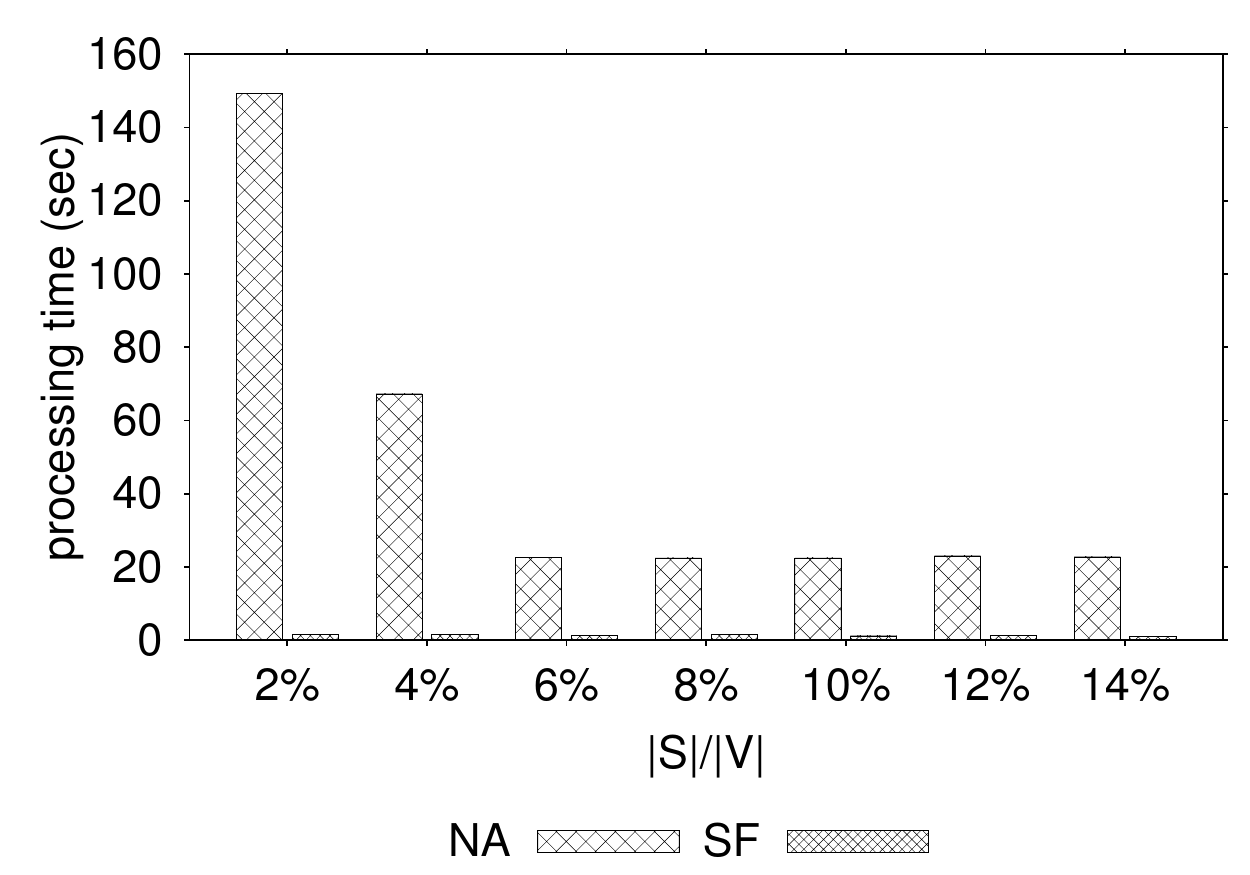}
    \label{fig:results:query:ssize}
    }
    \vspace{-5pt}
    \caption{Varying query parameters.} 
    \label{fig:results:query}
\end{figure}

To begin with, in Figure \ref{fig:results:system:alpha}, we illustrate 
how performance scales with regard to the smoothing factor $\alpha$ 
as it varies in $[0,1]$. Interestingly, we observe that execution times 
do not follow the same pattern. The reasons for this phenomenon lie 
with the different nature of each dataset. In particular, NA corresponds 
to the network of interstates in North America, whereas SF for the road 
network of San Francisco. Hence, the former is much sparser and the 
distances between the vertices are significantly greater while the 
fan-out degree is also significantly larger. The latter corresponds 
to a much denser network. Apparently, smoothing the edges over the 
population clusters has a more beneficial effect. Evidently the edge 
weights for partitioning the graph is more important as specific long 
edges server better so as to separate the different partitions and 
need to be used as cross-edges that isolate the vertices that are 
located closer to each other, so as to minimize the processing cost 
at the higher levels for merging the different partial results. 

In Figure \ref{fig:results:system:parallelism}, we present how execution 
time scales with the degree of parallelism. Apparently, there is an immense 
gain as the number of threads increases, a benefit that gradually fades 
for larger values. For the maximum gain that we observe for the NA dataset, 
just 22.5\% of the initial time is required for 8 threads, while for the SF 
dataset 34\% for the same number of threads. We reckon that this is a drastic 
improvement in the execution times. Another important observation is that our 
policy of having one global threshold which we can access only through acquiring 
the appropriate locks, does not pay off for a small number of partitions. In 
particular for a very small number of partitions, it would be more preferable 
if each thread had just one local distance threshold that does not share with 
any other thread as each partition is being processed, as if each partition was 
processed in isolation and in a later stage those results were combined accordingly. 
The overhead of having a subsequent is not dramatic since we propose this kind 
of processing for just 2 or 3 partitions, whereas for more partitions the hierarchical 
scheduling policy that we propose for processing the different partitions is ideal. 
The reason is that for just a few large partitions a separate distance threshold 
can be close enough to the distance value of the $k$-th element of any partial 
answer-set without the overhead that accompanies a global variable (e.g. waiting 
to acquire a lock, etc.). However, since we are interested in having small search 
areas, we see the balance leaning towards a global distance threshold for a greater 
number of partitions.


In Figure \ref{fig:results:query:kappa} we have a chance to study how 
the size of the result affects performance. Remarkably we observe two 
contradicting patterns. For SF we see that as the expected size of the 
result grows the required time increases. Interestingly though, the 
exactly opposite phenomenon is observed for NA, as we notice the 
execution time diminishing with $k$. Evidently, we can attribute this 
to the difficulty in building a result with the best pairs for low 
$k$ values, as with numerous local results we constantly have to read 
and update in a synchronized fashion the global distance threshold 
appropriately, given the default setting of partitions and threads. 
Clearly, there is an extremely high level of concurrency for that critical 
section of code where the anwer-set is augmented with new tuples for 
low $k$-values that causes a noticeable overhead as the threads have to 
wait so as to acquire the respective read and write locks. On the other 
hand, for the bigger by approximately 50,000 edges road network of San 
Francisco (SF), we have larger partitions (their number remains the same, 
but their size increases) and even though more effort is required within 
each separate partition, concurrency works best given the very different 
nature of the dataset: a densely populated area where the fan-out degree 
is significantly higher and the vertices have short distances from their 
neighbors. This is naturally reflected accordingly on the performance of 
the method. 

In Figure \ref{fig:results:query:rsize}, we illustrate the effect of the 
cardinality of the left operand of the join operation, while in Figure 
\ref{fig:results:query:ssize} the effect of the right operand. Evidently, 
there is a trade-off as these parameters increase. More specifically, 
it becomes easier to create a result-set of $k$ elements, and hence, 
performance seems to ameliorate with this parameter. However, for even 
greater values the processing cost dominates on performance and this is 
illustrated accordingly with an increase in the required time as $|R|$ 
and $|S|$ grow even further.

\newpage
\section{Conclusions}
\label{sec:concl}

To recapitulate, in this paper we presented a paradigm for 
processing in parallel graph joins in road networks. In 
particular, we address an otherwise extremely computationally 
expensive operation in the context of road networks. The 
methodology we propose matches the elements of two disjoint 
sets of nodes from the road network, with one preceding the 
other, say we want to visit a restaurant before a bar. We 
hence retrieve in parallel using concurrent mechanisms all 
eligible pairs of vertices $(u,v)$, where $u$ is a restaurant 
and $v$ is a bar, such that $\textrm{dist} (u,v)\leq \theta$, 
with $\theta$ the distance predicate, e.g., a walking distance. 
A variation of the problem would involve retrieving the $k$ 
closest pairs $(u,v)$, such that $\textrm{dist}(u,v) \leq 
\textrm{dist}(w,y)$, where $w,y$ do not belong in the result.
Moreover, we make use of a variety of parameters in order to 
tweak the partitioning scheme and we study their effect in our 
experimental evaluation. Finally, we relied on real-world data 
and showed how vulnerable and affected execution time is by the 
data distribution, and the way skewness can be mitigated using 
an appropriate system configuration.
\bibliography{joins}
\bibliographystyle{abbrv}
\balance
\end{document}